%% file: paper.tex
\documentclass[a4paper,11pth]{article}
\usepackage[T1]{fontenc}
\usepackage[utf8]{inputenc}
\usepackage{natbib}
\usepackage{amsmath,amssymb,amsfonts,amsthm}
\usepackage{booktabs}
\usepackage{graphicx}
\DeclareGraphicsExtensions{.pdf,.png,.jpg,.eps}
\usepackage{array}
\usepackage{multirow}
\usepackage{tabularx}
\usepackage{hyperref}
\newcolumntype{L}{>{\raggedright\arraybackslash}X}
\newcommand{\mv}[1]{{\boldsymbol{#1}}}

\title{Nowcasting Covid-19 statistics reported with delay: a case-study of Sweden}
\date{\today}
\author{
	Adam Altmejd\thanks{SOFI, Stockholm University and Swedish House of Finance, \href{mailto:adam@altmejd.se}{adam@altmejd.se}} \and
	Joacim Rocklöv\thanks{Department of Public Health and Clinical Medicine, Umeå University,\href{mailto:joacim.rocklov@umu.se}{joacim.rocklov@umu.se }}\thanks{Heidelberg Institute of Global Health, University of Heidelberg}  \and
	Jonas Wallin\thanks{Department of statistics, Lund university, \href{mailto:jonas.wallin81@gmail.com}{jonas.wallin81@gmail.com}}
}

\begin{document}
    \maketitle
    \begin{abstract}
        The new corona virus disease --- COVID-2019 --- is rapidly spreading through the world. The availability of unbiased timely statistics of trends in disease events are a key to effective responses. But due to reporting delays, the most recently reported numbers are frequently underestimating of the total number of infections, hospitalizations and deaths creating an illusion of a downward trend. Here we describe a statistical methodology for predicting true daily quantities and their uncertainty, estimated using historical reporting delays. The methodology takes into account the observed distribution pattern of the lag. It is derived from the ``removal method'', a well-established estimation framework in the field of ecology.
	\end{abstract}
\section{Pandemic response demands timely data}
The new corona virus pandemic is affecting societies all around the world. As countries are challenged to control and fight back, they are in need of timely, unbiased, data for monitoring trends and making fast and well-informed decisions \citep{No_author_2020_coronavirus_three}. Official statistics are usually reported with long delay after thorough verification, but in the midst of a deadly pandemic, real time data is of critical importance for policymakers \citep{Jajosky2004_evaluation_reporting}. The latest data are often not finalized, but change as new information is reported. In fact, reporting delays make the most recent days have the least cases accounted for, producing a dangerous illusion of an always improving outlook.

Still, these unfinished statistics offer crucial information. If the pandemic is indeed slowing, we should not wait for the data to be finalized before using it. Rather, we argue that actual case counts and deaths should be nowcasted to account for reporting delay, thus allowing policymakers to use the latest numbers availiable without beinig misled by reporting bias.

Such predictions provide an additional feature that is perhaps even more important. They explicitly model the uncertainty about these unknown quantities, ensuring that all users of these data have the same view of the current state of the epidemic.

In this paper we describe a statistical methodology for nowcasting the epidemic statistics, such as hospitalizations or deaths, and their degrees of uncertainty, based on the daily reported event frequency and the observed distribution pattern of reporting delays. The prediction model is building on methods developed in ecology, referred to as the ``removal method'' \citep{Pollock1991_review_papers}.

To help motivate why such forecasting is needed, we now turn to the case of Sweden. The model is flexible by design, however, and could easily be applied to other countries as well.

\subsection{The current situation of COVID-19 in Sweden}
The Swedish Public Health Agency updates the COVID-19 statistics daily\footnote{The data is published on \url{https://www.folkhalsomyndigheten.se/smittskydd-beredskap/utbrott/aktuella-utbrott/covid-19/bekraftade-fall-i-sverige/}.}. During a press conference, they present updates on the number of deaths, admissions to hospitals and intensive care, as well as case counts.

One of the reasons for following these indicators is to enable public health professionals and the public to observe the evolving patterns of the epidemic \citep{Anderson2020_how_will}. In relation to policy, it is of specific interest to understand if the growth rates changes, which could indicate the need for a policy response. However, in each daily report only a proportion of the number of recent deaths is yet known, and this bias produces the illusion of a downward trend.

The death counts suffer from the longest reporting delay. In their daily press conference, the Swedish Public Health Agency warns for this by stopping the reported 7-day moving average trend line 10 days before the latest date. But not only are deaths often reported far further back than 10 days, a bar plot still shows the latest information, creating a sense of a downward trend. In fact, this might be the reason why the number of daily deaths have been underestimated repeatedly. At the peak, deaths were initially believed to level out at around 60 per day, but after all cases had been reported more than two weeks later, the actual number was close to 120 \citep{Ohman2020_antalet_virusdoda}.

\section{The removal method}
We propose to use the removal method, developed in animal management \citep{Pollock1991_review_papers}, to present an estimate of the actual frequencies at a given day and their uncertainty. The method has a long history dating back at least to the 1930s \citep{Leslie1939_attempt_determine}. However, the first refined mathematical treatment of the method is credited to \citet{Moran1951_mathematical_theory}, more modern derivatives exits today \citep{Matechou2016_open_models}. It is a commonly applied method today when analyzing age cohorts in fishery and wildlife management.

The removal method that has three major advantages over simply reporting moving averages:

\begin{itemize}
	\item it does not relay any previous trend in the data,
	\item we can generate prediction intervals for the uncertainty about daily true frequencies,
	\item the uncertainty estimates can be carried over to epidemiological models to help create more realistic models.
\end{itemize}

A classic example where the method proposed to solve this problem has been used is in estimating statistics of trapping a closed population of animals \citep{Pollock1991_review_papers}. Each day the trapped animals are collected, and kept, and if there is no immigration the number of trapped animals the following days will, on average, decline. This pattern of declining number of trapped animals allows one to draw inference of the underlying population size. Here we replace the animal population with the true number of deaths
on a given day. Instead of traps we have the new reports of COVID--19 events. As the number of new reported deaths for a given day declines, we can draw inference on how many actually died that day. If we assume that the reporting structure is constant over time we can after a while quickly get good estimate of the actual number.

Suppose for example that on day one, 4 individuals are reported dead for that day. On the second day, 10 deaths are recorded for day two. Then, with no further information, it is reasonable to assume that more people died on day two. If the proportion reported on the first day is 3\%, the actual number of deaths would be 133 for day one and 333 for day two.

If additionally, 60 deaths are reported during the second day to have happened during day one, and on the third day, only 40 are reported for day two, we now have conflicting information. From the first-day reports it seemed like more people had died during day two, but the second day-reports gave the opposite indication. The model we propose systematically deals with such data, and handles many other sources of systematic variation in reporting delay. In fact, the Swedish reporting lag follows a calendar pattern. The number of events reported during weekends is much smaller. To account for this, we allow the estimated proportions of daily reported cases to follow a probability distribution taking into consideration what type of day it is.

\section{Applying the model to COVID-19 in Sweden}
We propose a Bayesian version of the removal model that assumes an overdispersed binomial distribution for the daily observations of deaths in Sweden in COVID-19. We then calculate the posterior distribution, prediction median and 95\% prediction intervals of the expected deaths from the reported deaths on each specific day. The method and algorithm is thoroughly described in the Supplementary Information.

To get accurate estimates we apply two institution-specific corrections. First, we only count workdays as constituting reporting delay, as very few deaths are reported during weekends. Second, we apply a constant bias correction to account for the fact that Swedish deaths come from two distinct populations with different trends: deaths in hospitals, and in elderly care.

\begin{figure}
    \centering
    \includegraphics[width=0.9\textwidth]{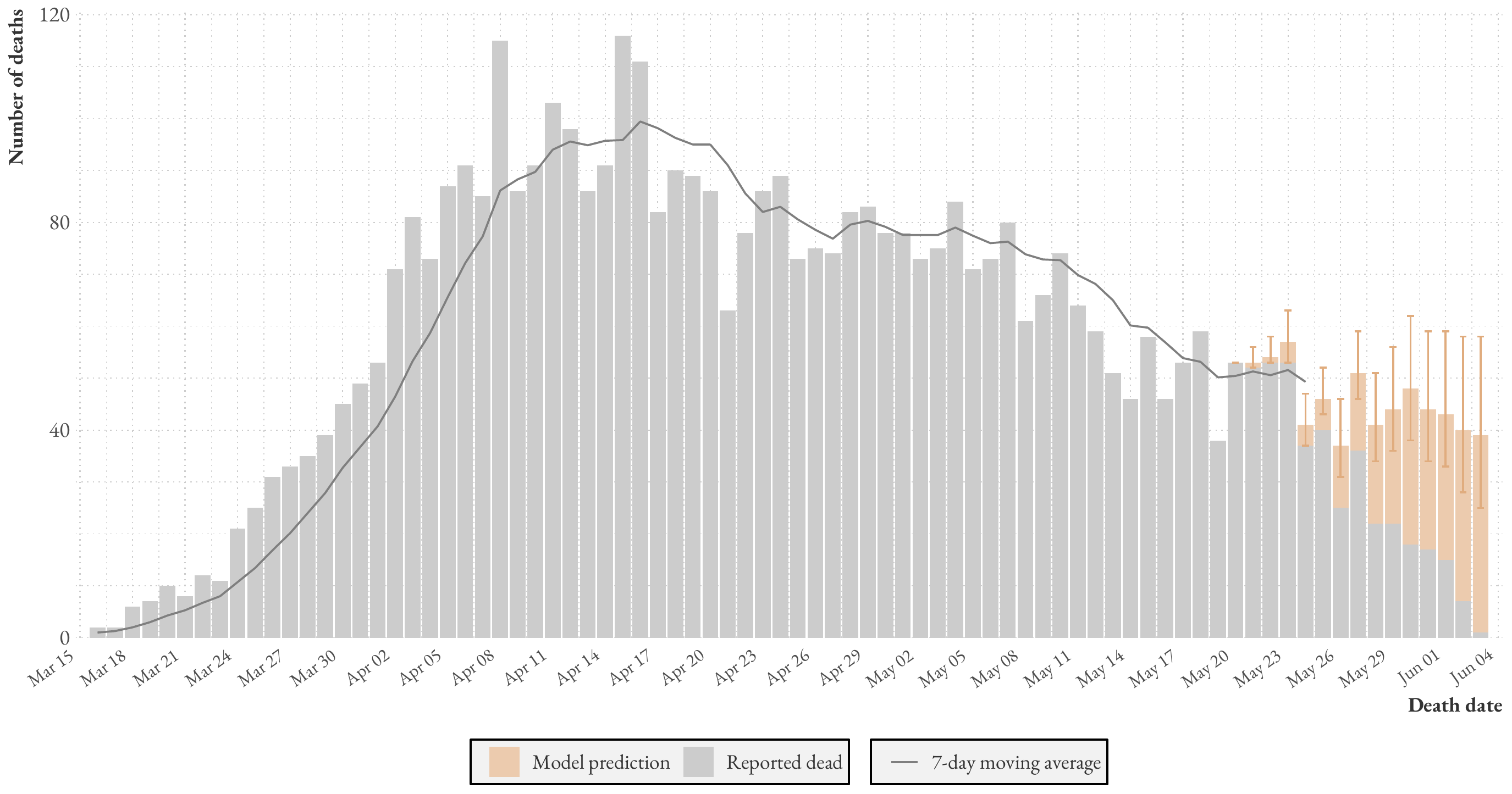}
    \caption{Swedish Covid-19 deaths as of 2020-06-04 and model predictions}
    \label{fig:latest_prediction}
\end{figure}

In Figure~\ref{fig:latest_prediction} we apply the model to the latest statistics from Sweden. The graph shows reported and predicted deaths (with uncertainty intervals) as bars, and a dashed line plots the 7-day (centered) moving average. A version without predictions is used in the Public Health Agency\'s daily press briefings. As expected, the model provides estimates of actual deaths considerably above the reported number of deaths. Not how the model predicts additional deaths above the moving average line.

\subsection{Model Performance}
To judge whether or not the model is accurate we need to compare it to a benchmark. The moving average of reported deaths is not useful, since it is biased for deaths that occurred within the last week. Instead, we create a benchmark prediction by a Normal distribution where the mean and standard deviation is taken from the historical lags from the last two weeks to the reported numbers\footnote{For a death date two days ago we add the mean of deaths reported after 3 days, 4 days, etc. We use the sum of standard deviations to generate the prediction intervals, assuming that lags are independent across days. The exact calculation is described in the appendix.}. 

Figure~\ref{fig:four_dates} depicts four randomly chosen dates where the model is compared to the benchmark. The model and the benchmark are tasked with predicting the total number of individuals who have died at a given date and have been reported within 14 days of that date. As time progresses, more deaths are reported and the dashed grey line approaches the horizontal line. Meanwhile model uncertainty decreases.

\begin{figure}
    \centering
    \includegraphics[width=0.9\textwidth]{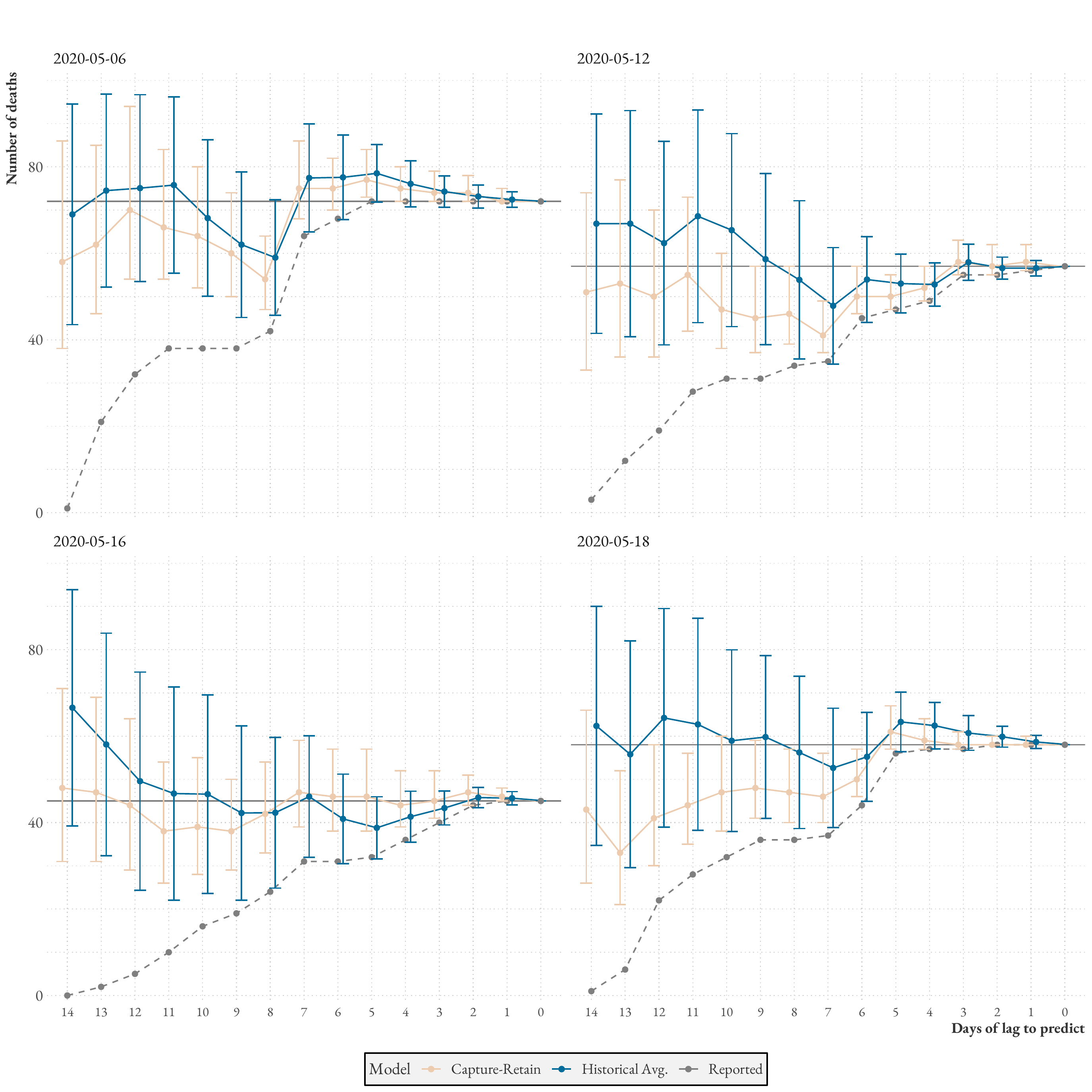}
    \caption{Model accuracy over time for four dates, compared to benchmark}
    \label{fig:four_dates}
\end{figure}

\begin{figure}
    \centering
    \includegraphics[width=0.9\textwidth]{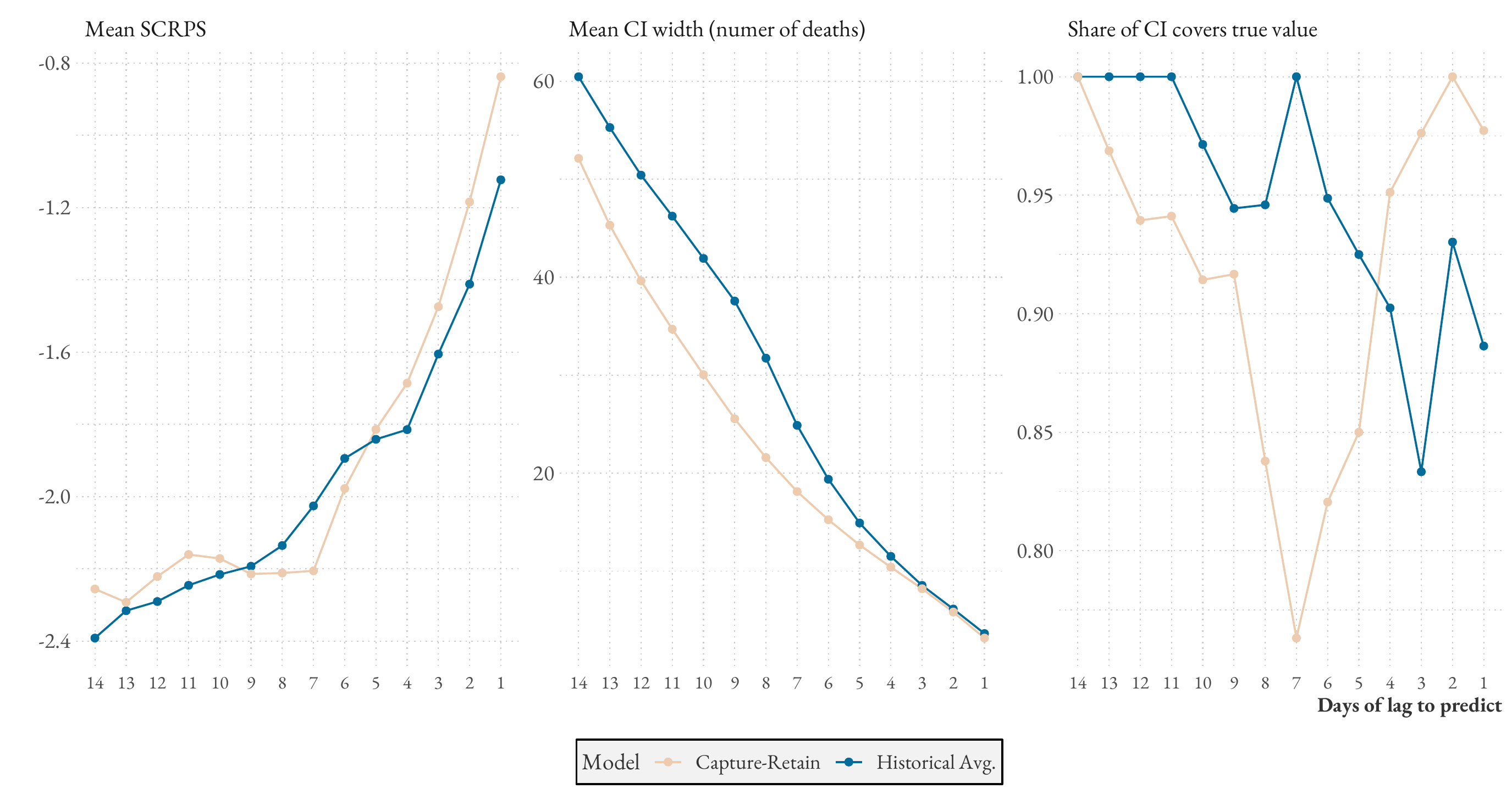}
    \caption{Model accuracy over time for four dates, compared to a benchmark}
    \label{fig:model_metrics}
\end{figure}

Figure~\ref{fig:model_metrics} shows model performance compared to the benchmark for three difference performance metrics. All three graphs are based on predictions of reported deaths within 14 days, and show how performance increases as more data has been reported. Each data point is the average of all dates where predictions can be evaluated. SCRPS is a measure of accuracy that rewards precision, it is a proper scoring rule like the continuous probability rank score or the Brier score (see definition in Appendix) \citep{bolin2019scale}. The central plot shows the width of the prediction intervals, and the rightmost one the proportion of PI\'s that cover the true value.

Benchmark and model point estimates are similarly close to the truth. The model produces tighter prediction intervals. For 8-5 days of reporting lag (see Figure \ref{fig:model_metrics}), the intervals are too tight. This is likely because the Public Health Agency queries the Swedish death registry for Covid-19 deaths only once or twice a week. Since we do not know the process, it has not been explicitly modeled.

\section{Implications and limitations}
The model proposed here can estimate the trends in surveillance data with reporting delays, such as the daily COVID-19 reports in Sweden. To generate accurate estimates of the actual event frequencies based on these reports is highly relevant and can have large implications for interpretations of the trends and evolution of disease outbreaks. In Sweden, delays are considerable and exhibit a weekday and holiday pattern that need to be accounted for to draw conclusions from the data. The method and algorithm proposed overcomes major shortcomings in the daily interpretation and practice analyzing and controlling the novel Corona virus pandemic. It also provides valuable measures of uncertainty around these estimates, showing users how large the range of possible outcomes can be.

Whenever case statistics are collected from multiple sources and attributed to its actual event date in the middle of a public health emergency, similar reporting delays to the ones in Sweden will necessarily occur. The method described thus has implications and value beyond Sweden, for any situation where nowcasts of disease event frequencies are of relevance to public health.

Nevertheless, the method also has its limitations. As presented, the model assumes that all deaths are reported in the same manner. Given there exists many regions in Sweden this is unlikely to be the case. For example, it is easy to see that the Swedish region Västra Götland follows a different reporting structure than Stockholm. Building a model for each region separately would most likely give better results and make the assumptions more reasonable. Unfortunately we do not currently have access to the high resolution data required to do so. 

Moreover, deaths are reported from two distinct populations that seem to follow different trends. At the time of writing, the daily deaths in elderly care, reported with a longer delay, seem to be decreasing slower than hospital deaths. But statistics offer only aggregate numbers, prohibiting us from modeling two distinct processes. However, we have noted a clear decline in proportions of deaths reported the two first working days. For example the number of deaths occurring at the second of April $\approx 30\%$ of deaths where reported within the first two working days whereas for the eighteens of May only $\approx 10\%$ where reported during the two first working days. We address this by assuming that the deaths reported during the two first working days comes from a different population then the remainder of days.

Another limitation is that the model assumes that the number of new reported deaths for a given day cannot be negative, which is not actually true, due to miscount or misclassification of days. The number of such cases is very small, however, and its removal should not make much difference. The central assumption of the model is that the proportions deaths reported each day is fixed (up to the known covariates). If actual reporting standards change over time, the model will not be able to account for this. But reporting likely becomes faster as the crisis infrastructure improves. One can imagine that after a while the reporting improves, or is changed, if this is not accounted for by a covariate in the model, it will report incorrect numbers. Of course, there might be unknown variables that we have failed to incorporate, but at the least the model is an improvement from the estimates using moving averages. When the covariates to the reporting delay pattern are known, the model can incorporate them and provide more accurate predictions. 

\section{Conclusion}
In this paper, we provide a method to accurately nowcast daily Covid-19 statistics that are reported with delay. By systematically modelling the delay, policy makers can avoid dangerous illusory downward trends. Our model also gives precise uncertainty intervals, making sure users of these statistics are aware of the fast-paced changes that are possible during this pandemic.

\bibliographystyle{chicago}
\bibliography{bibliography}
\appendix
\section{Appendix}
\input{appendix.tex}%

\end{document}

%% file: appendix.tex
\section{Model}
\subsection{Notation}
Before presenting the model we describe some notation used through out the appendix. For a $m \times n$ matrix $r$ we use the following broadcasting notation $\mv{r}_{k,j:l}=[ r_{k,j}, r_{k,j+1}, \ldots, r_{k,l}]$.
Further $x | y \sim \pi(.)$ implies that the random variable $x$ if we conditioning on $y$ follows distribution $\pi(.)$.
The relevant variables in the model are the following:

	\begin{tabularx}{\linewidth}{ccL}
		Variable name & Dimension & Description \\  \hline
		$\mv{d}$ & $T \times 1$ & $d_i$ is the number of deaths that occurred day $i$. \\
		$\mv{r}$ & $T \times T$ & $r_{ij}$ is number of death recorded for day $i$ at day $j$.  Note that $r_{ij}$ for $i<j$ is not defined.   \\
		$\mv{p}$ & $T \times T$ & $p_{ij}$ is the probability of that a death for day $i$ not yet recorded is recorded at day $j$.
		  Note that $p_{ij}$ for $i<j$ is not defined.  \\
		$\mv{\alpha}$ & $K \times 1$ & Latent prior parameter for $\mv{p}$ \\
		$\mv{\beta}$ & $K \times 1$ & Latent prior parameter for $\mv{p}$ \\
		$\mv{\alpha}^H$ & $2 \times 1$ & parameter for the probability, $\mv{p}$ for holiday adjustment. \\
		$\mv{\beta}^H$ & $2 \times 1$ & parameter for the probability, $\mv{p}$ for holiday adjustment. \\
		$\mv{\mu}$ &  $T \times 1$ &  $\mu_i$  is the intensity of the expected number of deaths at day $i$. \\
		$\sigma^2$ & $1\times 1$ & Variation of the random walk prior for the log intensity. \\
		$\phi$ & $1\times 1$ & overdispersion parameter for negative binomial distribution. \\
		$p_0$ & $1\times 1$ & probability of reporting for a low reporting event. \\
		$pi$ & $1\times 1$ & probability of a low reporting event.
	\end{tabularx}
\subsection{likelihood}
The most complex part of our model is the likelihood, i.e. the density of the observations given the parameters. Here the data consist the daily report of recorded deaths for the past days. This can conveniently be represented upper triangular matrix, $\mv{r}$, where $r_{i,j}$ represents number of new reported deaths for day $i$ reported at day $j$. This matrix is displayed on the left in Table \ref{tab:Data}.

\begin{table}
	\centering
	\begin{tabular}{cccccc}
		\multicolumn{1}{c}{} & \multicolumn{5}{c}{Reported date}                                             \\
		\parbox[t]{2mm}{\multirow{5}{*}{\rotatebox[origin=c]{90}{Death date}}}   & $r_{11}$ & $r_{12}$ & $\cdots$ &$\cdots$  &  $r_{1T}$\\
		& & $r_{22}$ &  $\cdots$ & $\cdots$   &$r_{2T}$ \\
		& & &$r_{33}$ &  $\cdots$ &  $r_{3T}$ \\
		& & & &  $\ddots$ & $\vdots$  \\
		& & & &  &  $r_{TT}$ \\

	\end{tabular}

	\caption{The table describes the observations data.}
	\label{tab:Data}
\end{table}

 We assume that given the true number of deaths at day $i$, $d_i$, that each reported day $j$ the remaining death $d_i - \sum_{k=1}^{j-1}r_{i,k}$ each recored with probability $p_{ij}$, i.e. $$r_{i,j}|D_i,r_{1,1:j}.p \sim Bin(d_i - \sum_{k=1}^{j-1}r_{i,k}, p_{i,j}).$$

Typically in removal sampling one would set the probability of reporting uniform, i.e. $p_{i,j}:=p$. However for this data this is clearly not realistic given weekly patterns in reporting --very little reporting during the weekends. Instead we assume that we have $k$ different probabilities. Further, to account for overdispertion, we assume that each probability rather being a fixed scalar is a random variable with a Beta distribution. The Beta distribution has two parameters $\alpha$ and $\beta$. This resulting the following distribution for the probabilities
$$
p_{i,j}| \mv{\alpha},\mv{\beta}, \mv{\alpha}^H,\mv{\beta}^H  \sim Beta(\alpha^H_j\alpha_{min(j-i,k)},\beta^H_j\beta_{min(j-i,k)}).
$$
Here, if $j\in H$ then day $j$ is a holidays or weekends, and the parameters above are
$$
\alpha^H_j = \begin{cases}
\alpha_1^H \alpha_2^H & \mbox{if }  \{j\in H \}\cup  \{j-1\in H \},  \\
\alpha_1^H & \mbox{if }  \{j\in H \}\cup  \{j-1\in H^c \}, \\
\alpha_2^H & \mbox{if }  \{j\in H^c \}\cup  \{j-1\in H \}, \\
1 & \mbox{else,}
\end{cases}
$$
and
$$
\beta^H_j = \begin{cases}
\beta_1^H \beta_2^H & \mbox{if }  \{j\in H \}\cup  \{j-1\in H \},  \\
\beta_1^H & \mbox{if }  \{j\in H \}\cup  \{j-1\in H^c \}, \\
\beta_2^H & \mbox{if }  \{j\in H^c \}\cup  \{j-1\in H \}, \\
1 & \mbox{else.}
\end{cases}
$$
These extra parameters are created to account for the under-reporting that occurs during weekend and holidays.
Finally we add an extra mixture component that allows for very low reporting.

\subsection{Priors}
For the $\mv{\alpha}$ and $\mv{\beta}$ parameters we use an (improper) uniform prior. For the deaths, $\mv{d}$, one could imagine several different prior ideally some sort of epidemiological model. However, here we just assume a log-Gaussian Cox processes \citep{Moller1998_log_gaussian}, but instead of Poisson distribution we use a negative binomial to handle possible over dispersion. The latent Gaussian processes has a intrinsic random walk distribution \citep{Rue2005_gaussian_markov} i.e.
\begin{align*}
\log(\mu_i) - \log(\mu_{i-1}) &\sim N(0,\sigma^2),\\
d_i| \lambda_i  &\sim NegBin(\mu_i, \phi).
\end{align*}
This model is created to create a temporal smoothing between the reported deaths.
For the hyperparameter $\sigma^2$ we impose a inverse Gamma distribution, this prior is suitable here because it guarantees that the process is not constant ($\sigma^2=0$) which we know is not the case.
\subsection{Full model}
Putting the likelihood and priors together we get the following hierarchical Bayesian model
\begin{align*}
\sigma^2 &\sim \Gamma(1,0.01) \\
\phi &\sim \Gamma(1,0.01) \\
\alpha_k &\sim U[0,\infty] \\
\beta_k &\sim U[0,\infty] \\
\alpha_k^H &\sim U[0,\infty] \\
\beta_k^H &\sim U[0,\infty] \\
\log(\mu_i) - \log(\mu{i-1}) &\sim N(0,\sigma^2)\\
d_i| \lambda_i  &\sim NegBin(\mu_i,\phi) \\
p_{i,j}|  \mv{\alpha},\mv{\beta}, \mv{\alpha}^H,\mv{\beta}^H &\sim Beta(\alpha^H_j\alpha_{min(j-i,k)},\beta^H_j\beta_{min(j-i,k)}) \\
r_{i,j}|d_i,\mv{r}_{1,1:j},p &\sim \pi Bin(d_i - \sum_{k=1}^{j-1}r_{i,k},p_0)+ (1-\pi) Bin(d_i - \sum_{k=1}^{j-1}r_{i,k}, p_{i,j}),
\end{align*}
where where and $j\leq i$ and $i=1,\ldots,T$.

\section{Inference}
As the main goal to generate inference of the number of death $\mv{d}$ is through the posterior distribution of number of deaths $\mv{d}$ given the observations $\mv{r}$.
In order to generate samples from this distribution we use a Markov Chain Monte Carlo method \citep{Brooks2011_handbook_markov}. In more detail we use a blocked Gibbs sampler, which generates samples in the following sequence:
\begin{itemize}
	\item  We sample $\mv{\alpha},\mv{\beta}, \mv{\alpha}^H,\mv{\beta}^H|\mv{d}, \mv{r}$ using the fact that one can integrate out $p$ in the model, and then  $\mv{d}|\mv{\alpha},\mv{\beta}, \mv{\alpha}^H,\mv{\beta}^H,\mv{r},\mv{\lambda}$  follows a Beta-Binomial distribution. Here to we use an adaptive MALA \citep{Atchade2006_adaptive_version} to sample from these parameters.
	\item  To sample $\mv{d}|\mv{\alpha},\mv{\beta}, \mv{\alpha}^H,\mv{\beta}^H,\mv{r},\mv{\lambda}$, that each death, $d_i$ is conditionally independent, and we just use a Metropolis Hastings random walk to sample each one.
	\item To sample $\mv{\lambda} | \mv{d},\sigma^2$ we again use an adaptive MALA.
	\item Finally We sample $\sigma^2|\mv{d}$,and  $p_0,\pi$ directly since this distribution is explicit, and $\phi$ using a MH-RW.
\end{itemize}

\section{Model Benchmark}
In this section, we present additional comparison of the model to the benchmark. We first describe the benchmark model in detail.

The benchmark model simply takes the sum of average historical reporting lags for the preceding 14 days. As before $r_{ij}$ is the number of deaths that happened on day $i$ and were recorded on day $j$. To predict the number of people that died on a given day, we first calculate lag averages:

\begin{align}
    \hat{r}_{i, i+L} = \frac{\sum^{14}_{k=i-14} r_{k - L, k}}{14},
\end{align}

where $\hat{r}_{i, i+L}$ is the average number of deaths reported with a lag of $L$ days, based on the 14 reports closest preceding day $i$. If we are looking at data released $2020-04-28$ and call this day 0, the latest death date that we have 10-day ($L=10$) reporting lag observation for is $r_{-10,0}$. The average for $Lag(0, 10)$ is therefore taken over the 14 days between $r_{-24,-14}$ and $r_{-10,0}$ (2020-04-04 and 2020-04-18). For this reason, some of the earlier predictions will not have data from $14$ days. The average is then taken over all available reports.

In the comparisons we aim at predicting the total number of deaths that will have been reported within 14 days of the death date. To do so, we sum over the average lag that has yet to be reported. If we are predicting the number of people that have yet to be reported dead for day -3, we already know the true values for $r_{-3,-3}$, $r_{-3,-2}$, $r_{-3,-1}$, and $r_{-3,0}$ so we only need to predict $r_{-3,1}\ldots r_{-3,10}$. The prediction is then

\begin{align}
    Benchmark(i, j) = \sum_{l=i}^{j} r_{i,l}+ \sum_{l = j}^{14} \hat{r}_{i, l}.
\end{align}

As confidence interval we simply use a Normal assumption with standard deviations of the reporting lags, assuming independence, i.e. this is just the square root of the sum of $Var(\hat{r})$.

\begin{figure}
    \centering
    \includegraphics[width=0.9\textwidth]{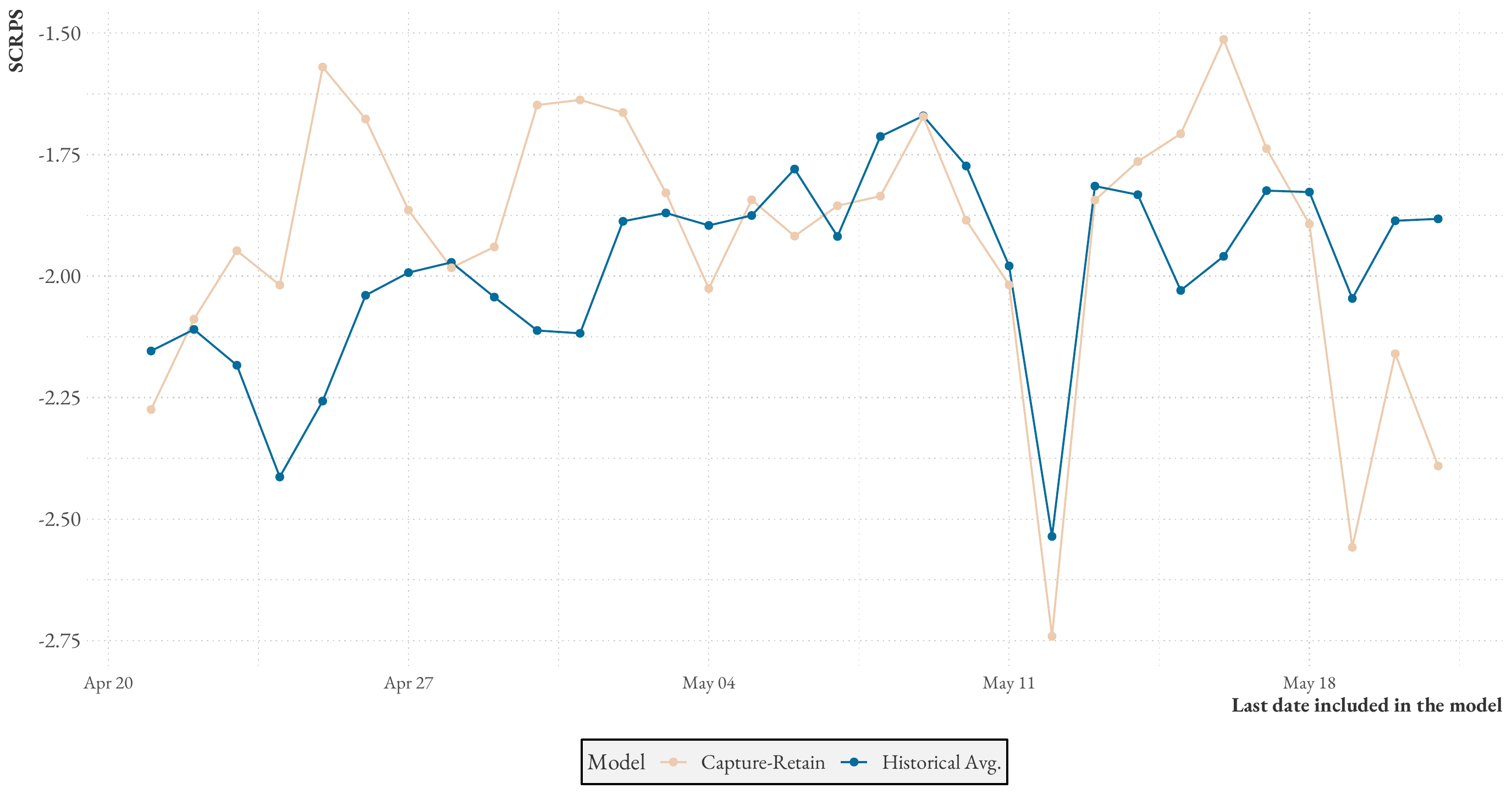}
    \caption{Average SCRPS as the pandemic progresses.}
    \label{fig:SCRPS_states}
\end{figure}

\begin{figure}
    \centering
    \includegraphics[width=0.9\textwidth]{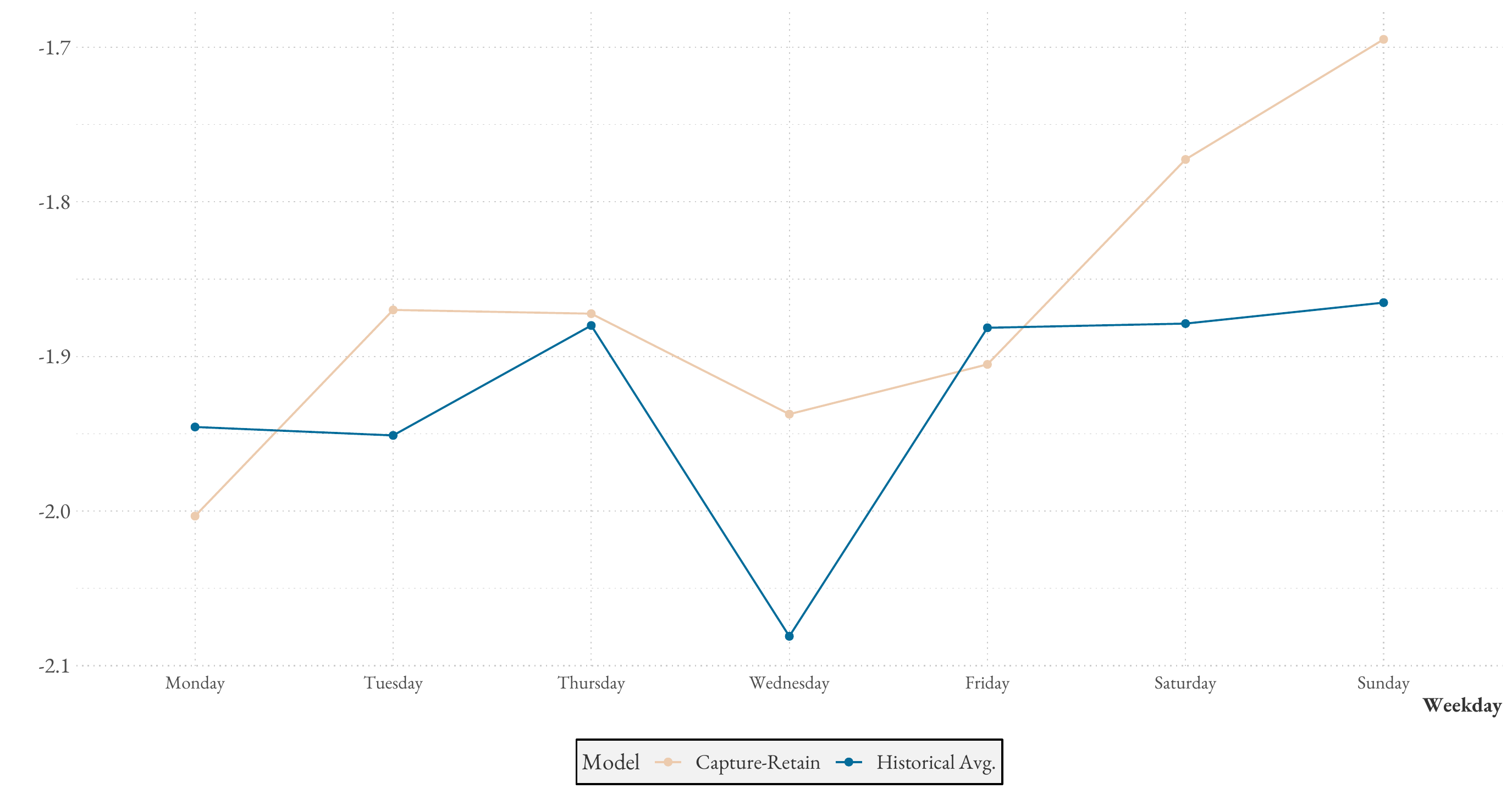}
    \caption{Average SCRPS per weekday.}
    \label{fig:SCRPS_weekdays}
\end{figure}

%% file: paper.bbl
\begin{thebibliography}{}

\bibitem[\protect\citeauthoryear{Anderson, Heesterbeek, Klinkenberg, and
  Hollingsworth}{Anderson et~al.}{2020}]{Anderson2020_how_will}
Anderson, R.~M., H.~Heesterbeek, D.~Klinkenberg, and T.~D. Hollingsworth
  (2020).
\newblock How will country-based mitigation measures influence the course of
  the {{COVID}}-19 epidemic?
\newblock {\em The Lancet\/}~{\em 395\/}(10228), 931--934.

\bibitem[\protect\citeauthoryear{Atchad{\'e}}{Atchad{\'e}}{2006}]{Atchade2006_adaptive_version}
Atchad{\'e}, Y.~F. (2006).
\newblock An {{Adaptive Version}} for the {{Metropolis Adjusted Langevin
  Algorithm}} with a {{Truncated Drift}}.
\newblock {\em Methodology and Computing in Applied Probability\/}~{\em
  8\/}(2), 235--254.

\bibitem[\protect\citeauthoryear{Bolin and Wallin}{Bolin and
  Wallin}{2019}]{bolin2019scale}
Bolin, D. and J.~Wallin (2019).
\newblock Scale dependence: Why the average crps often is inappropriate for
  ranking probabilistic forecasts.

\bibitem[\protect\citeauthoryear{Brooks, Gelman, Jones, and Meng}{Brooks
  et~al.}{2011}]{Brooks2011_handbook_markov}
Brooks, S., A.~Gelman, G.~Jones, and X.-L. Meng (2011).
\newblock {\em Handbook of {{Markov Chain Monte Carlo}}}.
\newblock {CRC Press}.

\bibitem[\protect\citeauthoryear{Jajosky and Groseclose}{Jajosky and
  Groseclose}{2004}]{Jajosky2004_evaluation_reporting}
Jajosky, R.~A. and S.~L. Groseclose (2004).
\newblock Evaluation of reporting timeliness of public health surveillance
  systems for infectious diseases.
\newblock {\em BMC Public Health\/}~{\em 4\/}(1), 29.

\bibitem[\protect\citeauthoryear{Leslie and Davis}{Leslie and
  Davis}{1939}]{Leslie1939_attempt_determine}
Leslie, P.~H. and D.~H.~S. Davis (1939).
\newblock An {{Attempt}} to {{Determine}} the {{Absolute Number}} of {{Rats}}
  on a {{Given Area}}.
\newblock {\em Journal of Animal Ecology\/}~{\em 8\/}(1), 94--113.

\bibitem[\protect\citeauthoryear{Matechou, McCrea, Morgan, Nash, and
  Griffiths}{Matechou et~al.}{2016}]{Matechou2016_open_models}
Matechou, E., R.~S. McCrea, B.~J.~T. Morgan, D.~J. Nash, and R.~A. Griffiths
  (2016).
\newblock Open models for removal data.
\newblock {\em Annals of Applied Statistics\/}~{\em 10\/}(3), 1572--1589.

\bibitem[\protect\citeauthoryear{M{\o}ller, Syversveen, and
  Waagepetersen}{M{\o}ller et~al.}{1998}]{Moller1998_log_gaussian}
M{\o}ller, J., A.~R. Syversveen, and R.~P. Waagepetersen (1998).
\newblock Log {{Gaussian Cox Processes}}.
\newblock {\em Scandinavian Journal of Statistics\/}~{\em 25\/}(3), 451--482.

\bibitem[\protect\citeauthoryear{Moran}{Moran}{1951}]{Moran1951_mathematical_theory}
Moran, P. A.~P. (1951).
\newblock A {{Mathematical Theory}} of {{Animal Trapping}}.
\newblock {\em Biometrika\/}~{\em 38\/}(3/4), 307--311.

\bibitem[\protect\citeauthoryear{Nature}{Nature}{2020}]{No_author_2020_coronavirus_three}
Nature (2020).
\newblock Coronavirus: Three things all governments and their science advisers
  must do now.
\newblock {\em Nature\/}~{\em 579\/}(7799), 319--320.

\bibitem[\protect\citeauthoryear{{\"O}hman and Gagliano}{{\"O}hman and
  Gagliano}{2020}]{Ohman2020_antalet_virusdoda}
{\"O}hman, D. and A.~Gagliano (2020).
\newblock Antalet virusd\"oda har underskattats.
\newblock {\em Sveriges Radio\/}.

\bibitem[\protect\citeauthoryear{Pollock}{Pollock}{1991}]{Pollock1991_review_papers}
Pollock, K.~H. (1991).
\newblock Review {{Papers}}: {{Modeling Capture}}, {{Recapture}}, and {{Removal
  Statistics}} for {{Estimation}} of {{Demographic Parameters}} for {{Fish}}
  and {{Wildlife Populations}}: {{Past}}, {{Present}}, and {{Future}}.
\newblock {\em Journal of the American Statistical Association\/}~{\em
  86\/}(413), 225--238.

\bibitem[\protect\citeauthoryear{Rue and Held}{Rue and
  Held}{2005}]{Rue2005_gaussian_markov}
Rue, H. and L.~Held (2005).
\newblock {\em Gaussian {{Markov Random Fields}}: {{Theory}} and
  {{Applications}}}.
\newblock {CRC Press}.

\end{thebibliography}
